# Lattice Boltzmann methodology for unconfined flows

**Vemu Sahiti[1*], PS Gurugubelli[2], VK Surasani[3]**

[1]Department of Mechanical Engineering, BITS-Hyderabad, *500078*, Telangana, India
[2]Department of Mechanical Engineering, BITS-Hyderabad, *500078*, Telangana, India
[3]Department of Chemical Engineering, BITS-Hyderabad, *500078*, Telangana, India
*p20200041@hyderabad.bits-pilani.ac.in

## Abstract

Numerical analysis of unconfined flow over an obstacle has always been challenging in computational fluid dynamics due to the truncation of the computational domain while replicating the real-life flows and the application of the boundary conditions. Confined flows studies have been well established and documented while unconfined flow studies are relatively challenging. Present work demonstrates the implementation of lattice Boltzmann method for unconfined flow over a circular cylinder for Re 100. The cylinder was placed at 10D upstream and 30D downstream and 10D from both the top and bottom walls. Different boundary conditions were implemented at the top and bottom walls to ensure unconfined flow. Drag and lift coefficients are also presented and were computed using the momentum exchange algorithm. Results are in complete agreement with the existing literature which demonstrate the capability of the solver.

## 1. Introduction

Traditional computational methods utilize the Navier-Stokes equations by discretizing them over the whole domain by using techniques like the finite volume method, finite element method etc. Instead of considering the domain at a continuum scale, there exists an intermediate level at which the flow is studied as a packet of particles and the probability of them having a particular range of velocities with respect to space and time. Over the past couple of decades, lattice Boltzmann method (LBM) has emerged as a potential approach for solving fluid flows by solving the evolution of the particle distribution functions at the mesoscopic level (Chen & Doolen, 1998), (Succi, 2001). Due to the absence of the non-linear advection term, LBM is easier when it comes to parallelization (Kandhai, Koponen, Hoekstra, Kataja, Timonen & Sloot, 1998).

Everyday real-life flows around bodies like aircraft wings, submarines and underwater vehicles etc. take place in infinitely large domains. However, implementing the numerical analysis of such problems force us to truncate the domain for obvious reasons like the computational costs. Numerical analysis is done by discretizing the governing equations and applying the boundary conditions at the interfaces. To solve the equations at the boundaries, one needs to know the information of the flow parameters which are true conditions at infinity but are unknown. This is addressed by placing the obstacle far enough from the boundaries so that the applied boundary conditions do not affect the solution. However, the way boundary treatment is done has an impact on the accuracy and performance of the solver.

The flow characteristics around a circular cylinder depends on the dimensionless number, Reynolds number (Re) which is defined as UD/ν exhibiting different flow regimes for different Re. For example, at Re less than 5, the flow remains unattached and steady. As the Re increases from 40 to 50, flow starts to become unsteady and the wake starts to become unstable and vortex shedding is observed (Roshko, 1961). At Re = 100, the flow is characterized by two-dimensional laminar vortex shedding with well-defined Strouhal number (St ≈ 0.164) and relatively stable force coefficients, making it an ideal validation case for numerical methods (Williamson & Brown, 1998).

For confined flows over a body, no-slip condition is a well-established boundary condition and there has been a lot of work carried out for confined flow over bodies. In traditional solvers, it is made sure that the velocity of the fluid along the wall remains the same as the velocity of the wall. It is implemented in LBM via simple bounce-back condition which ensures the penetration of fluid into the solid is avoided (Ladd, 1994). Also, due to the flipped direction of the particle distribution functions after streaming, the no-slip is inherently achieved with $u=0$ and $v=0$. However, implementing unconfined flow becomes tricky because of the truncation of the domain and replication of the boundary conditions and relatively less documentation has been done in the present literature. LBM implements boundary conditions by manipulating the fundamental mesoscopic parameter—the particle distribution functions (also called populations)—for specific discrete directions. This manipulation ensures that the macroscopic properties satisfy the desired boundary conditions. Present work demonstrates the implementation of these boundary conditions for flow past a circular cylinder at Re 100.

## 2. Lattice Boltzmann methodology

### 2.1 Outline of LBM

To describe the behavior of a fluid, one does not need to know the dynamics of every molecule or a particle but a collective statistical behavior of the particles can represent a system. Boltzmann equation is the equation which describes the statistical behavior of a system which is deviated from its equilibrium. Non-dimensional, continuous and force-free Boltzmann equation is written as,

$$\frac{\partial f}{\partial t} + \xi_\alpha \frac{\partial f}{\partial x_\alpha} = \Omega(f) \quad (1)$$

Traditionally, LBM is based on Cartesian grid with lattice structure such as D2Q9 and D3Q17. The development of grid in LBM was originally based on cartesian grid where D is the dimensions and Q is the number of quadrature points.

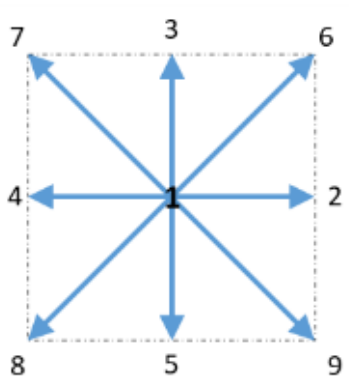

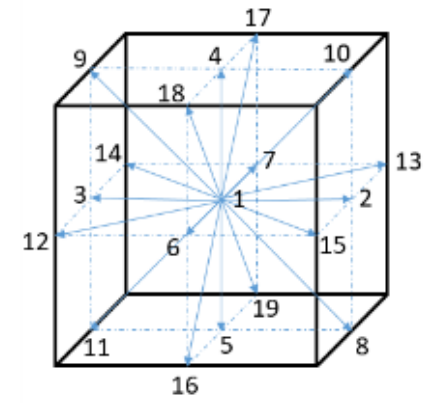

**Figure 1**: D2Q9 and D3Q17 lattice configurations for standard LBM lattice

LBM is straight forward method which has two main operations: collision and streaming. Force-free lattice Boltzmann equation with a uniform lattice and BGK collision model can be expressed as,

$$f_i(\vec{x} + \vec{c}_i \Delta t, t + \Delta t) = f_i(\vec{x}, t) - \frac{1}{\tau}[(f_i(\vec{x}, t) - f^{eq}(\vec{x}, t)] \quad (2)$$

where $f_i$ is the particle distribution function along the particle speed direction $\vec{c}_i$ at position $\vec{x}$ and time $t$. $f_i^{eq}$ is the equilibrium distribution function and the rate at which equilibrium is approached is governed by $\tau$. The equilibrium distribution function is of the form,

$$f_i^{eq} = \omega_i \rho \left[1 + \frac{3}{c^2}\vec{c}_i.\vec{u} + \frac{9}{2c^4}(\vec{c}_i.\vec{u})^2 - \frac{3}{2c^2}\vec{u}.\vec{u}\right] \quad (3)$$

where, $c = \Delta x/\Delta t$ is the lattice speed, and $\Delta x$ and $\Delta t$ are the lattice width and time step, respectively. Here, $\Delta t$ is chosen to be equal to $\Delta x$, thus, $c = 1$. $\omega_i$ is a weighing factor. The particle speed $\vec{c}_i$ and $\omega_i$ for D2Q9 lattice model (Fig.1) are represented as given in Table.1.

**Table 1**: Velocity sets and weights of D2Q9 configuration.

| ***i*** | **1** | **2** | **3** | **4** | **5** | **6** | **7** | **8** | **9** |
|---|---|---|---|---|---|---|---|---|---|
| $cx_i$ | 0 | 1 | 0 | -1 | 0 | 1 | -1 | -1 | 1 |
| $cy_i$ | 0 | 0 | 1 | 0 | -1 | 1 | 1 | -1 | -1 |
| $w_i$ | 4/9 | 1/9 | 1/9 | 1/9 | 1/9 | 1/36 | 1/36 | 1/36 | 1/36 |

After the computation of the particle distribution functions, the macroscopic properties are found from the moment equations as,

$$\rho = \sum_{i=1}^{n} f_i, \ \rho\vec{u} = \sum_{i=1}^{n} f_i \vec{c}_i \quad (4)$$

The viscosity and the relaxation parameter from the collision part of the LBE are related as,

$$\upsilon = c_s^2 \Delta t(\tau - 0.5) \quad (5)$$

From the above relation, as $\tau$ approaches 0.5, the viscosity approaches zero. This causes instability in the solver causing it to crash. To address that, we relax the different directions of the pdfs with different moments with the help of MRT(multiple relaxation time) collision model proposed by (d'Humières, Ginzburg, Krafczyk, Lallemand & Luo, 2002). This model is implemented by using a transformation matrix to work with the macroscopic properties in the moment space. The collision operator here is

$$\Omega_\alpha = -\sum_i (M^{-1}SM)_{\alpha i}(f_\beta(\vec{x}, t) - f_i^{eq}(\vec{x}, t)) \quad (6)$$

where, S is the relaxation diagonal matrix which contains the relaxation rates of all the nine moments. Below is the relaxation used in the present study.

$$S = diag\left[1, 1.1\ , 1.1, 1, 1.1, 1, 1.1, \frac{1}{\tau}, \frac{1}{\tau}\right]$$

For D2Q9 model, the equilibria for the nine directions in the moment space are given as,

$$m_\alpha^{eq} = [\rho, e^{eq}, \varepsilon^{eq}, j_x^{eq}, q_x^{eq}, j_y^{eq}, q_y^{eq}, p_{xx}^{eq}, p_{xy}^{eq}]$$

$$m_\alpha^{eq} = \begin{bmatrix} 1 \\ -2 + 3(u^2 + v^2) \\ 1 - 3(u^2 + v^2) \\ u \\ -u \\ v \\ -v \\ u^2 - v^2 \\ uv \end{bmatrix}$$

The transformation matrix M is

$$M = \begin{bmatrix} 1 & 1 & 1 & 1 & 1 & 1 & 1 & 1 & 1 \\ -4 & -1 & -1 & -1 & -1 & 2 & 2 & 2 & 2 \\ 4 & -2 & -2 & -2 & -2 & 1 & 1 & 1 & 1 \\ 0 & 1 & 0 & -1 & 0 & 1 & -1 & -1 & 1 \\ 0 & -2 & 0 & 2 & 0 & 1 & -1 & -1 & 1 \\ 0 & 0 & 1 & 0 & -1 & 1 & 1 & -1 & -1 \\ 0 & 0 & -2 & 0 & 2 & 1 & 1 & -1 & -1 \\ 0 & 1 & -1 & 1 & -1 & 0 & 0 & 0 & 0 \\ 0 & 0 & 0 & 0 & 0 & 1 & -1 & 1 & -1 \end{bmatrix}$$

We perform the collision and then stream the post-collision populations to the neighboring lattices. After the streaming we implement the boundary conditions which in a way are the corrections of the populations which reflect the required conditions for the macroscopic properties.

### 2.2 Conversion factors

There has been a considerate amount of work in LBM. But there has always been a gap with the way the work was communicated. It is mostly presented in units relevant to mesoscales and it would be challenging for the person working in macroscale to comprehend the solution. To address that, there are conversion factors for length and time since they are the primitive units. $\Delta x^*$ and $\Delta t^*$ are the grid size and time-step size respectively in lattice scale and are taken as 1. In order to convert it into physical scale we select the conversion factor $\Delta x$ as 1/D. Present work addresses this gap with the inclusion of the mesoscale to physical scale conversion factors.

### 2.3 Boundary conditions

After streaming, the boundary conditions are applied to correct the populations which are missing or have incorrect values at the borders of the domain. For example, in Fig.2 it can be seen that at the top wall $f_5$, $f_8$, $f_9$ are the ones which need the corrections and similarly for the bottom wall, $f_3$, $f_6$, $f_7$ are the ones which need the corrections. Unconfined flow can be implemented in LBM via free-slip, periodic, non-equilibrium bounce-back, constant velocity bounce-back.

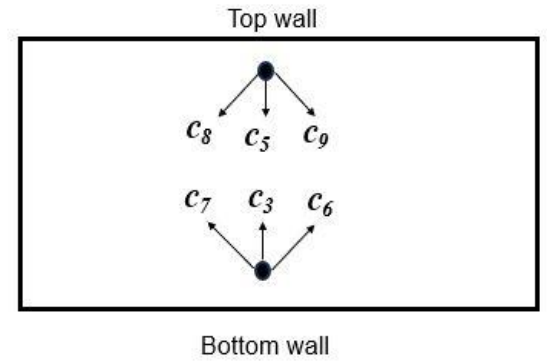

**Figure 2:** Missing populations at the end of the domain which stream from outside the domain.

#### 2.3.1 Free-slip

Free-slip condition (Sbragaglia & Succi, 2005) ensures that the normal velocity of the fluid is zero and the tangential part of the velocity is non-zero by specular reflection (Fig.3) of the particle distribution functions. In present work it is applied on the top and bottom walls and is applied after streaming.

$$f_{[5,8,9]}(top) = f_{[3,7,6]}(top) \quad (7a)$$

$$f_{[3,7,6]}(bottom) = f_{[5,8,9]}(bottom) \quad (7b)$$

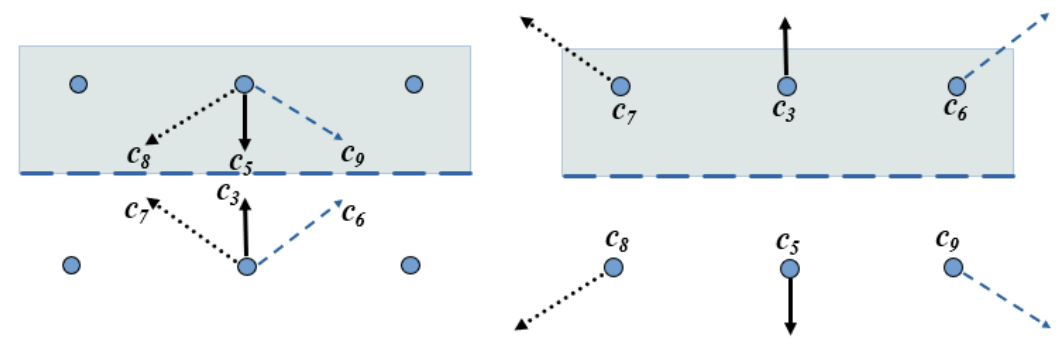

**Figure 3**: Free-slip boundary condition where the normal component of the velocity become zero and tangential is non-zero.

#### 2.3.2 Periodic

Periodic boundary conditions (PBC) are used when we want to simulate a spatially repeating system or eliminate the artificial effects of domain boundaries on the flow physics. This condition applies to the situation where the flow leaves the domain, enters back into the domain from the opposite side. In the present work it is implemented at the top and bottom walls and the correction is made to enforce the similar condition at both the walls. Periodic condition ensures that the mass is conserved at all times because mass leaving is mass entering (Succi, 2001).

$$f_{[5,8,9]}(top) = f_{[5,8,9]}(bottom) \quad (8a)$$

$$f_{[3,7,6]}(bottom) = f_{[3,7,6]}(top) \quad (8b)$$

#### 2.3.3 Non-equilibrium bounce-back (NEBB)

This condition was first proposed by Zou and He, (Zou & He, 1997) where the non-equilibrium part of the particle distribution function was bounced back and the missing populations were computed with a series of equations obtained from the moment equations. In the present work, it is implemented as follows. For top wall,

$$f_5 = f_3 \quad (9a)$$

$$f_8 = f_6 + 0.5(f_2 - f_4) - 0.5\rho u \quad (9b)$$

$$f_9 = f_7 - 0.5(f_2 - f_4) + 0.5\rho u \quad (9c)$$

For bottom wall,

$$f_3 = f_5 \quad (10a)$$

$$f_6 = f_8 - 0.5(f_2 - f_4) + 0.5\rho u \quad (10b)$$

$$f_7 = f_9 + 0.5(f_2 - f_4) - 0.5\rho u \quad (10c)$$

#### 2.3.4 Constant velocity bounce-back

Constant velocity bounce-back (Aidun, Lu & Ding, 1998), (Ladd, 1994) is implemented with an assumption that the top and bottom walls are moving with the same velocity as that at the inlet. From eqn.11a and 11b it can be seen that there is a term highlighted in violet color which accounts for the momentum transfer needed to maintain the prescribed velocity.

This ensures that the fluid adjacent to the boundary matches the wall's velocity, satisfying the no-slip condition for moving boundaries. It is also implemented after streaming and the corrections are done with post-collision populations. This is similar to the implementation of the bounce-back condition for a moving boundary.

$$f_{[5,8,9]}(top) = f^{star}_{[3,6,7]}(top) - 2w_{[3,6,7]}\rho(top)\frac{\vec{c}_{[3,6,7]}\cdot\vec{u}_w}{c_s^2} \quad (11a)$$

$$f_{[5,8,9]}(top) = f^{star}_{[3,6,7]}(top) - 2w_{[3,6,7]}\rho(top)\frac{\vec{c}_{[5,8,9]}\cdot\vec{u}_w}{c_s^2} \quad (11b)$$

#### 2.3.5 Stress-free outlet

A stress-free outlet represents a boundary where the fluid exits the boundary without experiencing any viscous stresses. In traditional numerical works it is implemented for a vertical outlet as,

$$\frac{\partial u}{\partial y} = 0;\ \frac{\partial v}{\partial y} = 0 \quad (12)$$

Present implementation is a practical and widely-used approximation to stress-free conditions. Present work it is implemented by zeroth order extrapolation. It is implemented as,

$$f_{outlet} = f_{outlet_adjacent} \quad (13)$$

### 2.4 Wet-node and link-wise approach

Treatment of the solid boundaries can be broadly classified into two approaches: wet-node and link wise. In wet-node approach, the node is simply either fluid node or solid node and the computational boundary is in between the nodes with alignment to the cartesian grid (Ladd, 1994), (Zou & He, 1997). This method is well suited for grid aligned boundaries like flat walls or square geometries. However, it can also be implemented for curved geometries which will be discussed in the next section. In the present study wet-node approach has been opted for cylindrical geometry by implementing simple bounce-back at the interface. Since the walls are aligned with the cartesian grid, wet-node implementation of the boundary conditions is implemented.

Link-wise approach applies the boundary conditions along the links of solid and fluid node which allows the boundary to cut through the links (Bouzidi, Firdaouss & Lallemand, 2001). Interpolation of the required functions allow one to capture the complex geometry, taking into account the distance of the boundary from the fluid node. The most common implementation is the interpolated bounce-back method, where the boundary intersection with each link is characterized by a fraction parameter (0 < q < 1) indicating the distance from the fluid node to the boundary. The reflected distribution function is then calculated using interpolation between the incoming distribution and equilibrium values, effectively placing the boundary at the correct geometric location rather than at grid nodes.

### 2.5 Momentum exchange algorithm

Forces at the cylinder are computed using the momentum exchange algorithm (Mei, Luo & Shyy, 1999). Complex boundaries approximate the computational boundary by stair-case approximation (Chen & Doolen, 1998) as shown in the fig.5. where it is assumed to be in between two grid spacings similar to simple bounce-back condition. From Fig.4 the yellow circles are the nodes which have the lattice links with the solid nodes (black circles). Red circles are the mid-way of the boundary nodes. Black dotted square is the computational boundary using the stair-case approximation. The momentum is calculated only for those populations which cross the boundary and bounce back at the solid node as shown in the figure. For flat boundaries (Fig.5), it is straight forward and for curved boundaries, every direction crossing the boundary is identified and momentum exchange is applied. It is calculated as,

$$\Delta \boldsymbol{P} = \sum_{\boldsymbol{all\ links}} \left(f_i^{in} + f_i^{out}\right)\boldsymbol{c_i} \quad (14)$$

Force component in x-direction is drag and y-direction is lift. Force coefficients are then obtained from,

$$c_d = \frac{F_x}{0.5\,\rho_{in}\,U_{in}^2\,D} \qquad c_l = \frac{F_y}{0.5\,\rho_{in}\,U_{in}^2\,D} \quad (15)$$

**Figure 4**: Stair-case approximation of curved boundaries while applying the bounce-back and computing the momentum exchange.

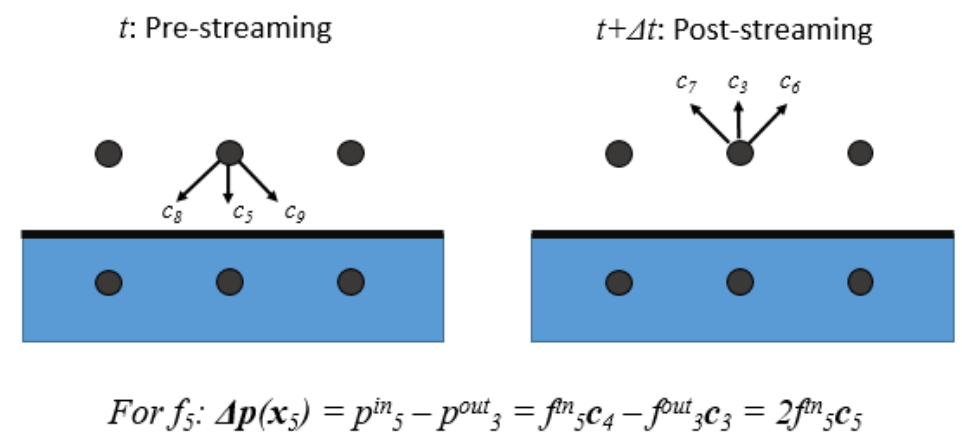

**Figure 5:** Momentum exchange algorithm for flat boundary. Post-collision $f^*_5, f^*_8, f^*_9$ after bouncing back become $f_3, f_6, f_7$ respectively.

## 3. Numerical setup

### 3.1 Flow past circular cylinder

Simulations have been carried out for unconfined flow past a circular cylinder and the domain specifications are presented in Fig.7. The cylinder's centre is placed at 10D upstream and 30D downstream. It is 10D from both top and bottom walls with D as 20 lattices. Inlet is subjected to uniform velocity of 0.02. Present study used a grid resolution of 800 and 400 in x and y directions with Re 100. Non-equilibrium bounce-back (Zou & He, 1997) is applied at the inlet to compute the unknown populations streaming into the domain and stress-free condition is applied at the outlet.

The same boundary conditions have been implemented for a finer grid in which D is taken as 40 lattices which makes a grid resolution of 1600 and 800 in x and y directions. Similar results as that of coarse grid are obtained and this shows that the solver is independent of the grid resolution. The results are not presented here due to space constraint. Besides, the main scope of the work is to make the reader understand the implementation of LBM and different boundary conditions along with selection of collision model, conversion factors and computation of the stability parameter.

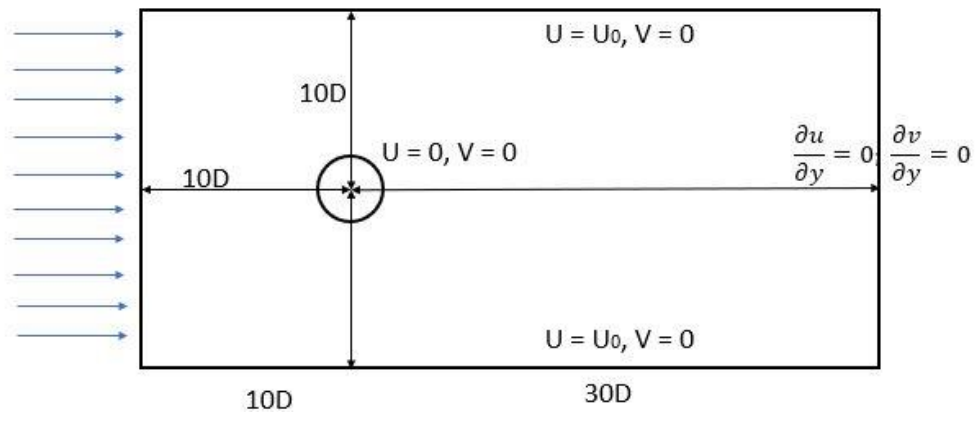

**Figure 6:** Schematic of the computational domain with uniform inlet, free-slip condition with tangential velocity at top wall being inlet velocity and normal velocity being zero.

The relaxation parameter, is computed from the Re as,

$$\nu = \frac{Du_{in}}{Re} \quad (16)$$

$$\tau = 3\nu + 0.5 \quad (17)$$

As τ gets close to 0.5, the stability of the solver becomes tricky. As the inlet velocity gets close to 0.1, the Mach number increases, flow field gets distorted and the stability of the solver becomes questionable. This in turn affects the evolution of the force coefficients' periodicity. So, the inlet velocity in the present work is selected as 0.02 and the MRT collision model is chosen because of low relaxation parameter.

Present study utilised the half-way bounce-back at the cylinder boundary where the populations are flipped back at the computational boundary and this happens in the same time step. Whereas, in full way bounce back, the populations reach the solid node one time step, and then they return to the fluid node in the next timestep.

In the present work, we have taken $\Delta x$ as 1/20 and $u_{physical}$ as 1. Inlet velocity $u^*$ is 0.02. The conversion is done according to the below equation and $\Delta t$ is $10^{-3}$. The value of $\Delta t$ is utilised in the calculation of Strouhal number for lift coefficient's periodicity.

$$\Delta t = \frac{\Delta x}{u_{physical}} u^* \quad (18)$$

### 3.2 Results and discussion

Simulations for unconfined flow past circular cylinder with different boundary conditions at Re 100 have been carried out. It has been observed that the force coefficients have not changed much with change in the boundary conditions at the walls. Below are the velocity magnitude contours of all the four boundary conditions. Periodic condition velocity contour could not be included due to space constraint but the contour is similar to the rest of the contours.

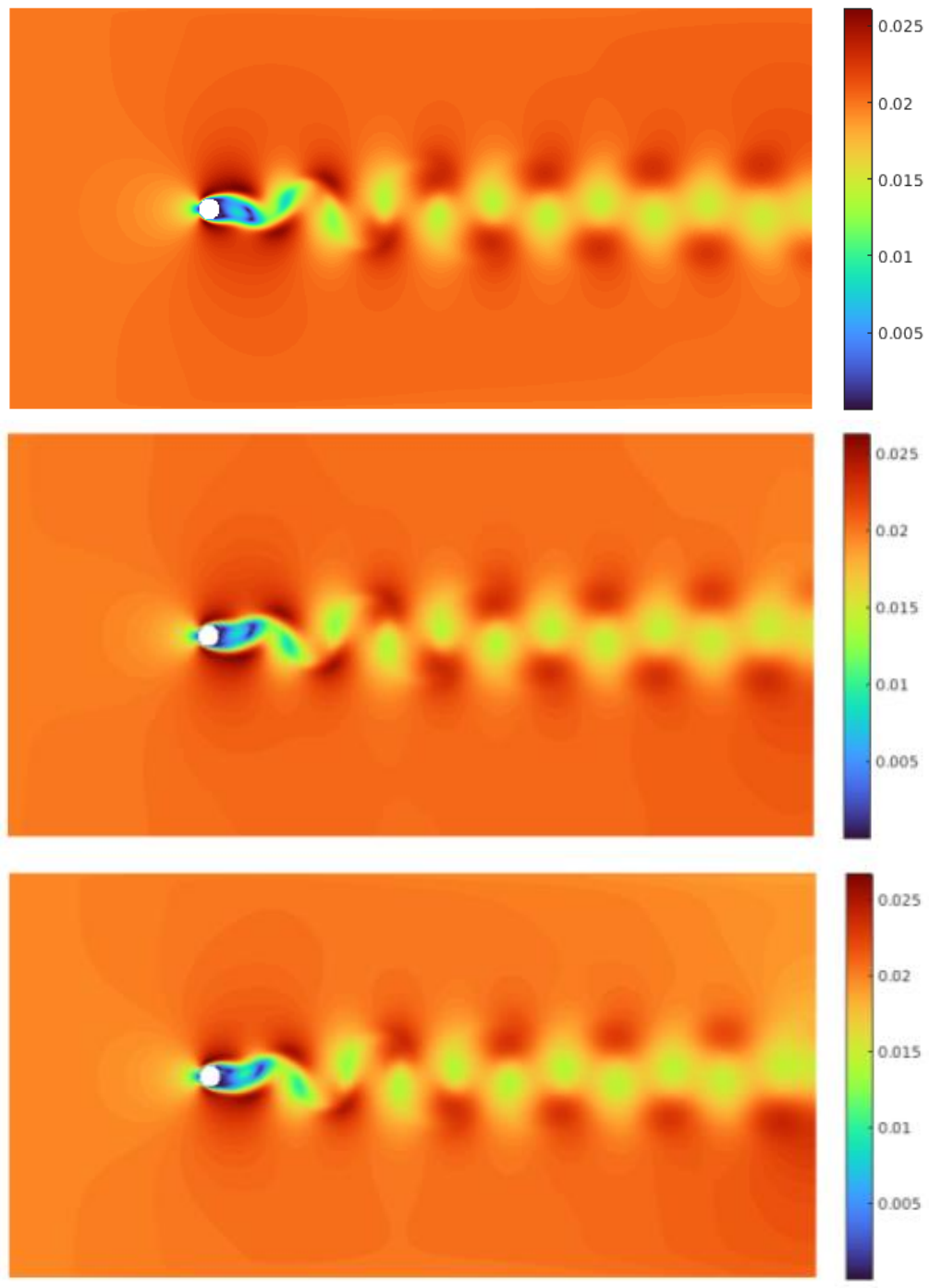

**Figure 7**: Velocity magnitude contours for unconfined flow past circular cylinder of different boundary conditions for Re 100

As it can be seen from the velocity magnitude contours, there is no impact of the boundary conditions on the flow field. The velocity contours are obtained from any LBM solver but it is the force coefficients which tell us the accuracy of the solver. There has been good amount of work on the qualitative analysis of LBM but very little work has been done on the quantitative aspect of it.

Below is the evolution of drag and lift coefficients with time for all the four boundary conditions. It can be seen that the force coefficients have not been altered much with change in the boundary conditions. From fig.9, it can be seen that the constant velocity bounce-back and NEBB boundary conditions are a slightly higher drag coefficient when compared to the free stream conditions (free-slip and periodic) which validate the conventional behaviour of traditional numerical approaches due to the formation of boundary layer at the walls. Comparison of the force coefficients with existing literature is also done and is presented below in Table 2.

**Table 2:** Drag and lift coefficients for unconfined flow past circular cylinder at constant Re 100.

| Work | Mean $C_d$ | $C_l$ | St |
|---|---|---|---|
| Braza, Chassaing, Ha Minh(1986) | 1.4 | 0.32 | 0.16 |
| Posdziech, Grundmann(2007) | 1.33 | 0.33 | 0.164 |
| Present work | 1.4 | 0.31 | 0.168 |

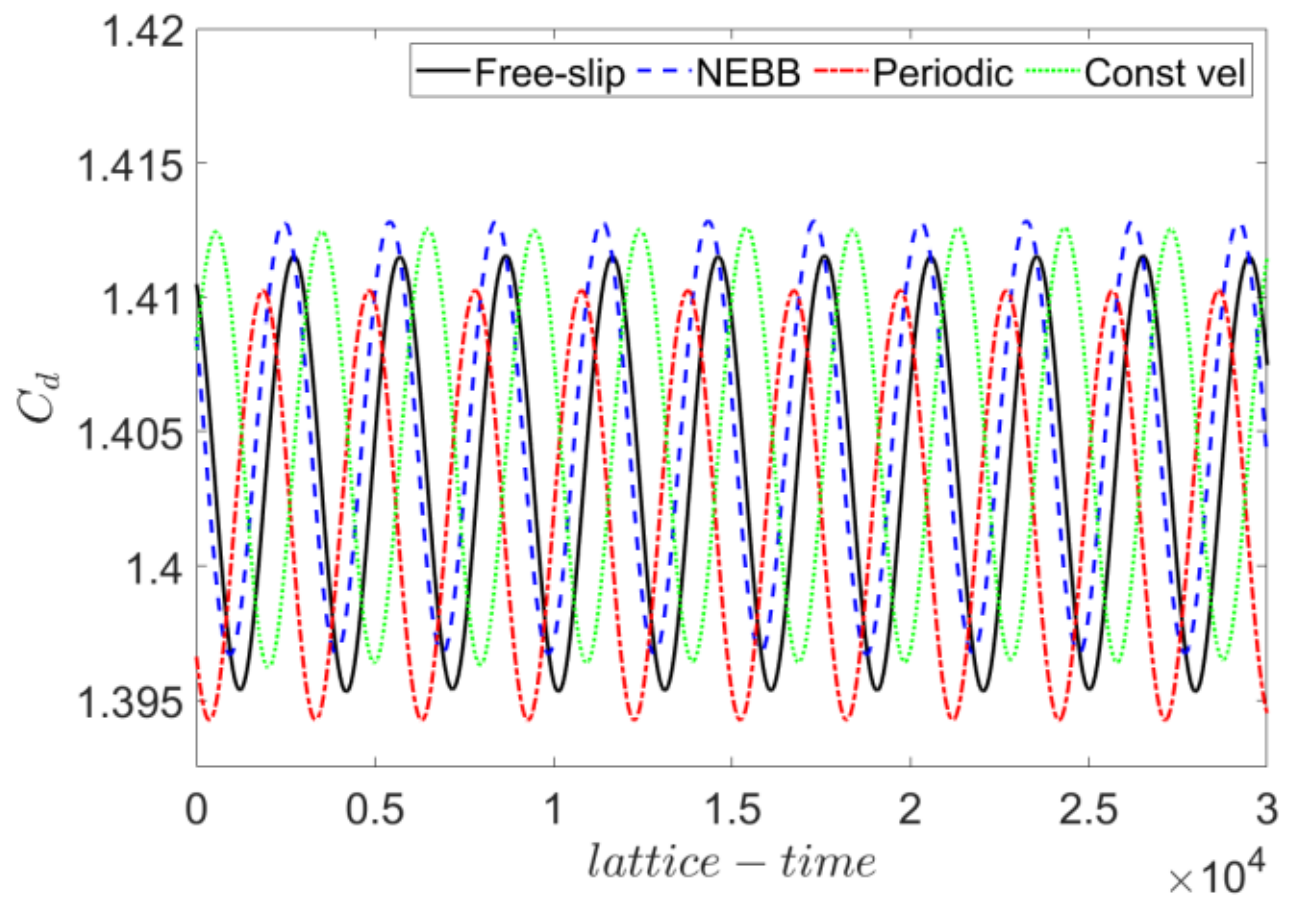

**Figure 8:** Drag coefficient for different boundary conditions at top and bottom walls for Re 100.

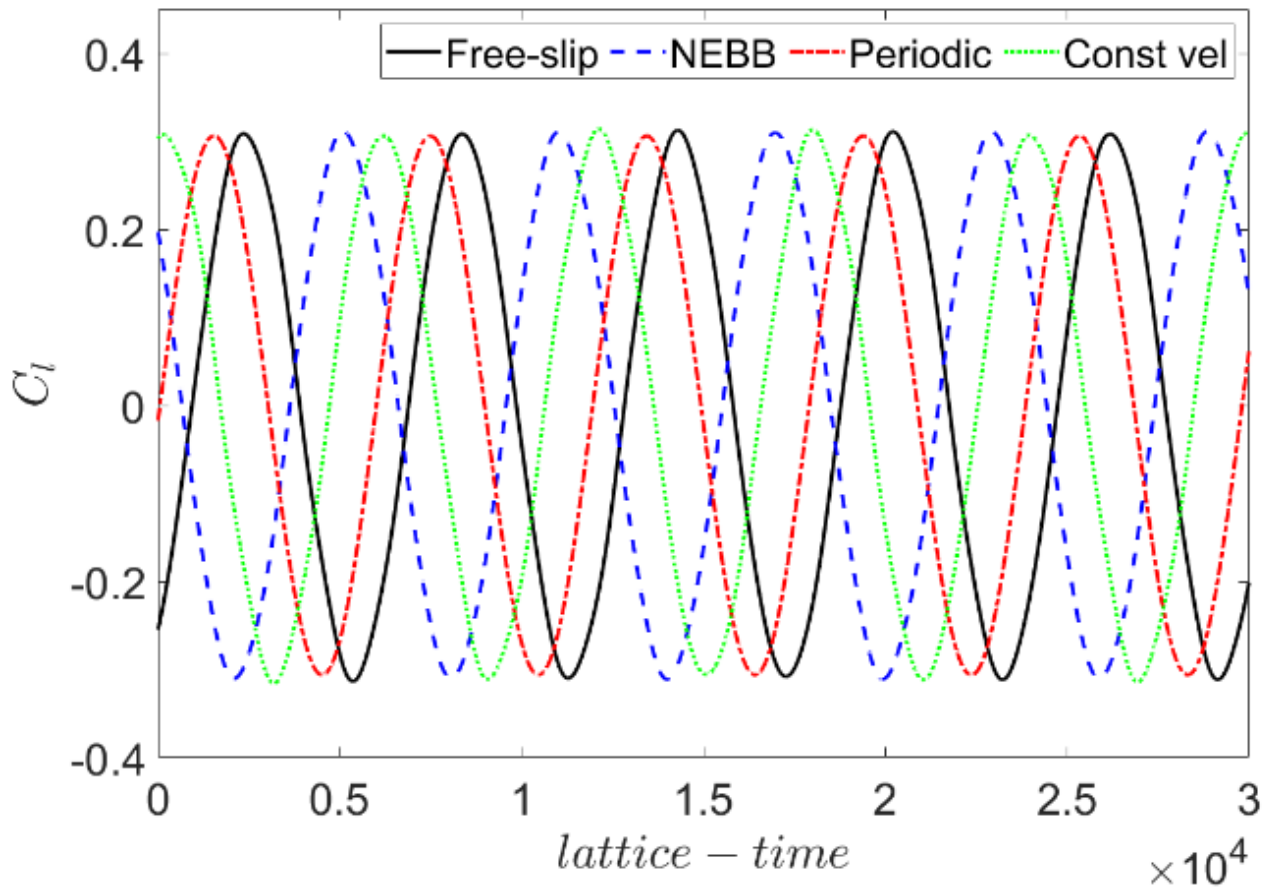

**Figure 9:** Lift coefficient for different boundary conditions at top and bottom walls for Re 100.

## 4. Conclusions

The lattice Boltzmann method has gained popularity over the past couple of decades due to its ease of implementation and computational advantages. A comprehensive lattice Boltzmann methodology has been implemented and presented for unconfined flow past a circular cylinder. Four different boundary conditions namely, free-slip, periodic, non-equilibrium bounce-back and constant velocity bounce-back have been applied at the top and bottom walls in order to achieve the conditions for the macroscopic properties.

From the drag coefficient plot, the $C_d$ value for NEBB and constant velocity bounce-back boundary conditions are slightly higher than the free stream boundary conditions. The boundary conditions did not affect the flow field and the lift coefficient. The results are in agreement with the existing literature which shows the accuracy of the solver. Since the force coefficients agree well with existing literature, LBM can be considered as an alternative approach to traditional conventional Navier-stokes solvers.